\documentclass[prl,twocolumn,showpacs,preprintnumbers,superscriptaddress]{revtex4-1}
\usepackage{amsmath}
\usepackage{amsfonts}
\usepackage{amssymb}
\usepackage{graphicx}% Include figure files
\usepackage{dcolumn}% Align table columns on decimal point
\usepackage{color}
\usepackage{bm}% bold math

\pdfoutput=1

\begin{document}

\title{Singular elastic strains and magnetoconductance of suspended graphene}

\author{E. Prada}
\author{P. San-Jose}
\affiliation{Department of Physics, Lancaster University, UK-LA1 4YB Lancaster,
United Kingdom}

\author{G. Le\'on}
\affiliation{Instituto de Ciencia de Materiales de Madrid, CSIC, Sor Juana In\'es de la Cruz 3, E28049 Madrid, Spain}

\author{M. M. Fogler}
\affiliation{Department of Physics, University of
California San Diego, 9500 Gilman Drive, La Jolla, California 92093, USA}

\author{F. Guinea}
\affiliation{Instituto de Ciencia de Materiales de Madrid, CSIC, Sor Juana In\'es de la Cruz 3, E28049 Madrid, Spain}

\begin{abstract}

Graphene membranes suspended off electric contacts or other rigid supports are prone to elastic strain, which is concentrated at the edges and corners of the samples. Such a strain leads to an algebraically varying effective magnetic field that can reach a few Tesla in sub-micron wide flakes. In the quantum Hall regime the interplay of the effective and the physical magnetic fields causes backscattering of the chiral edge channels, which can destroy the quantized conductance plateaus.

\end{abstract}


\maketitle

%%%%%%%%%%%%%%%%%%%%%%%%%%%%%%%%%%%%%%%%%%%%%%%%%%%%%%%%%%%%%%%%%%%
%{\em Introduction.}

The isolation of graphene monolayers~\cite{Netal04, Netal05} and the observation of the integer quantum Hall effect (QHE) in these systems~\cite{Netal05b, ZTSK05} have made graphene a very active research topic~\cite{NGPNG09, AGSci09}. The integer QHE in graphene is remarkably robust and can be seen even at room temperature~\cite{Netal07}. The traditional setup to measure the quantized Hall plateaus involves at least four contacts: source and drain for the current, and two (or more) side contacts for the voltage. This scheme helps to eliminate the spurious contact resistance. The much higher mobility of suspended samples~\cite{Betal08a, DSBA08, BSHSK08} raised hopes for observing also the fractional QHE in graphene. Yet demonstrating even the integer QHE in a four-contact setup proved to be difficult in such samples. Only recently the integer QHE has been confirmed in suspended graphene~\cite{DSBA08, Skachko2009iaf, Bolotin2009oot} by reverting to a two-contact scheme. (Similar observations have been made for bilayers~\cite{FMY09}.) The fractional QHE has also been reported in these experiments~\cite{Skachko2009iaf, Bolotin2009oot}. An artifact of the two-contact setup is the suppression of the quantized conductance~\cite{Skachko2009iaf, Bolotin2009oot}, familiar from the QHE in semiconductors~\cite{McEuen1990nrf, Richter1992ooq, FS94, Shlimak2004cow}. The reason why the nominally superior multi-contact scheme is less successful in suspended graphene has not been fully clarified, except for one theory that in small samples the side contacts had to be placed too close to the source and drain, causing admixture of the longitudinal and Hall conductances~\cite{Skachko2009iaf}.

%
% FIG. 1
%
\begin{figure}
\includegraphics[width=3.0in]{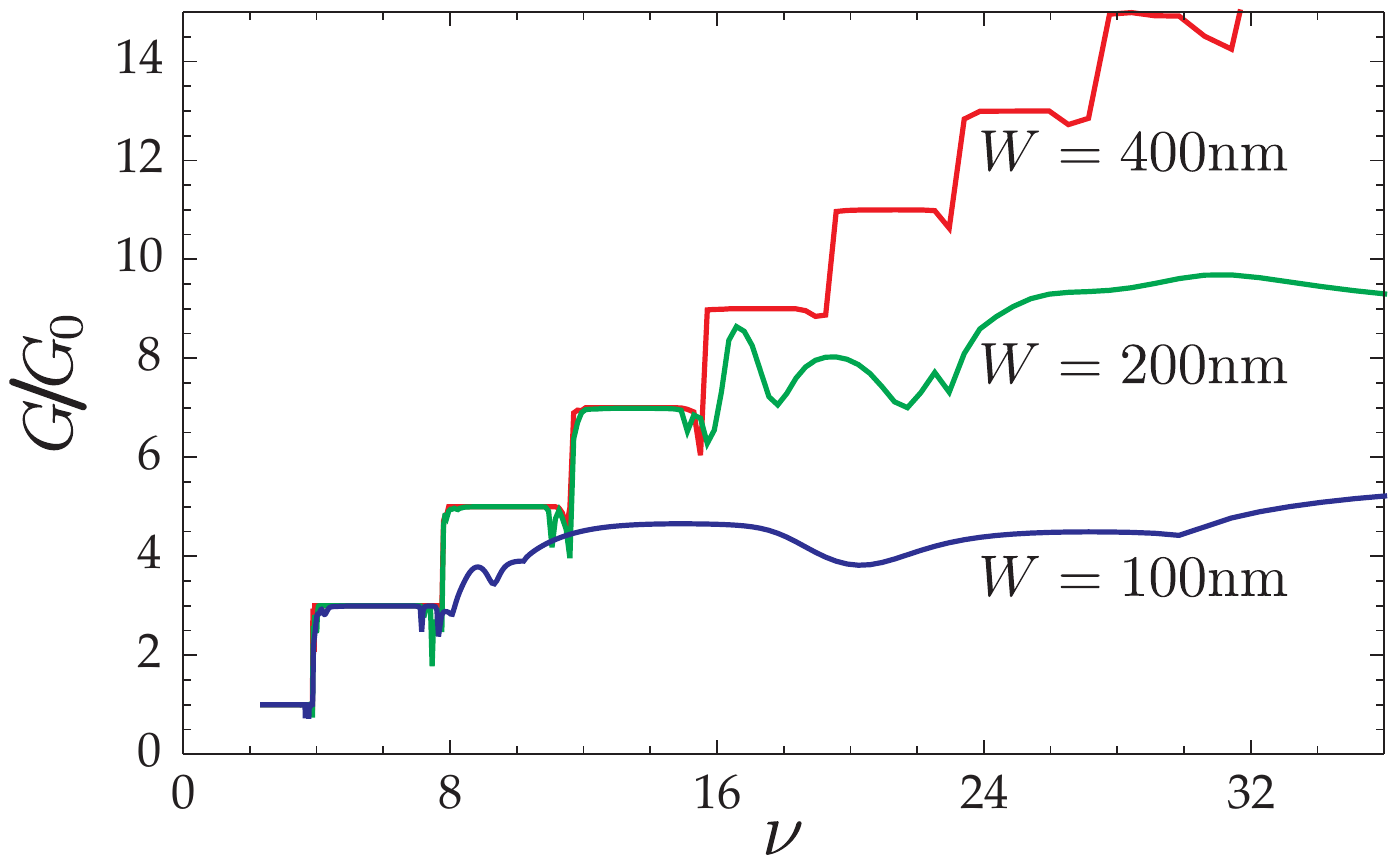}
\caption[fig]{Conductance in units of $G_0 = 2 e^2 / h$ as a function of the filling factor $\nu$ for samples of electron density $2.8 \times 10^{11}\,\text{cm}^{-2}$ and widths $W = 100$, $200$ and $400\,\text{nm}$. The side contacts are assumed to have width $2l = 30\,\text{nm}$ and to pull laterally outward with the force $P = 78\,\text{nN}$ ($c_0 = 1$) each.}
\label{fig:transmission}
\end{figure}

In this paper we consider a different effect, which may also contribute to the lack of quantization: when the graphene sheet is under tension, the side contacts induce a long-range elastic deformation which acts as a pseudomagnetic field $\mathbb{B}(x, y)$ for its massless charge carriers~\cite{SA02b, M07, NGPNG09}. The tension can be generated either by the electrostatic force of the underlying gate~\cite{FGK08, WPE09}, by interaction of graphene with the side walls~\cite{Betal08b}, or as a result of thermal expansion~\cite{Betal09, Tetal09}.
Our main results are as follows. We show that $\mathbb{B}$ is concentrated near the corners of the contacts where it exhibits power-law singularities. It decays into the interior of the sample but for a reasonable $0.1\%$ average strain in a 200-nm wide strip, $\mathbb{B}$ can remain of the order of a Tesla across its entire width. This leads to backscattering of the QHE edge states when the \emph{real} magnetic field $B$ is in a similar range, causing the erosion of the quantized conductance plateaus. We give an analytical argument that predicts that the QHE plateaus are destroyed above a threshold Landau level index $N_c$ and the corresponding filling factor $\nu_c$:
\begin{equation}
          N_c \sim 0.07 c_0^{-1/3} (k_F W)^{4/3}\,,
          \quad
          \nu_c = 4 N_c + 2\,,
\label{eqn:N_c}
\end{equation}
where $k_F$ is the Fermi momentum in zero magnetic field, $W$ is the width of the sample, and $c_0 \sim 1$ (see below). We also compute the conductance numerically. The results for $k_F = 0.94 \times 10^6\, \text{cm}^{-1}$ and three representative $W$'s are shown in Fig.~\ref{fig:transmission}. For these three traces Eq.~\eqref{eqn:N_c} gives (top to bottom) $\nu_c \approx 38$, $16$, and $7.6$, in agreement with the simulations. Based on these results, we suggest that observation of the QHE in suspended graphene with side contacts requires samples of width $W \geq 200\,\text{nm}$. On the other hand, in small samples one can envision purposely using the pseudomagnetic field as a new tool for tuning electronic properties of graphene nanostructures~\cite{PNP09, PN09, GKG09, FR09}.

%%%%%%%%%%%%%%%%%%%%%%%%%%%%%%%%%%%%%%%%%%%%%%%%%%%%%%%%%%%%%%%%%%%
%{\em The model. Two terminal setup.}

\begin{figure}
\includegraphics[width=2.8in]{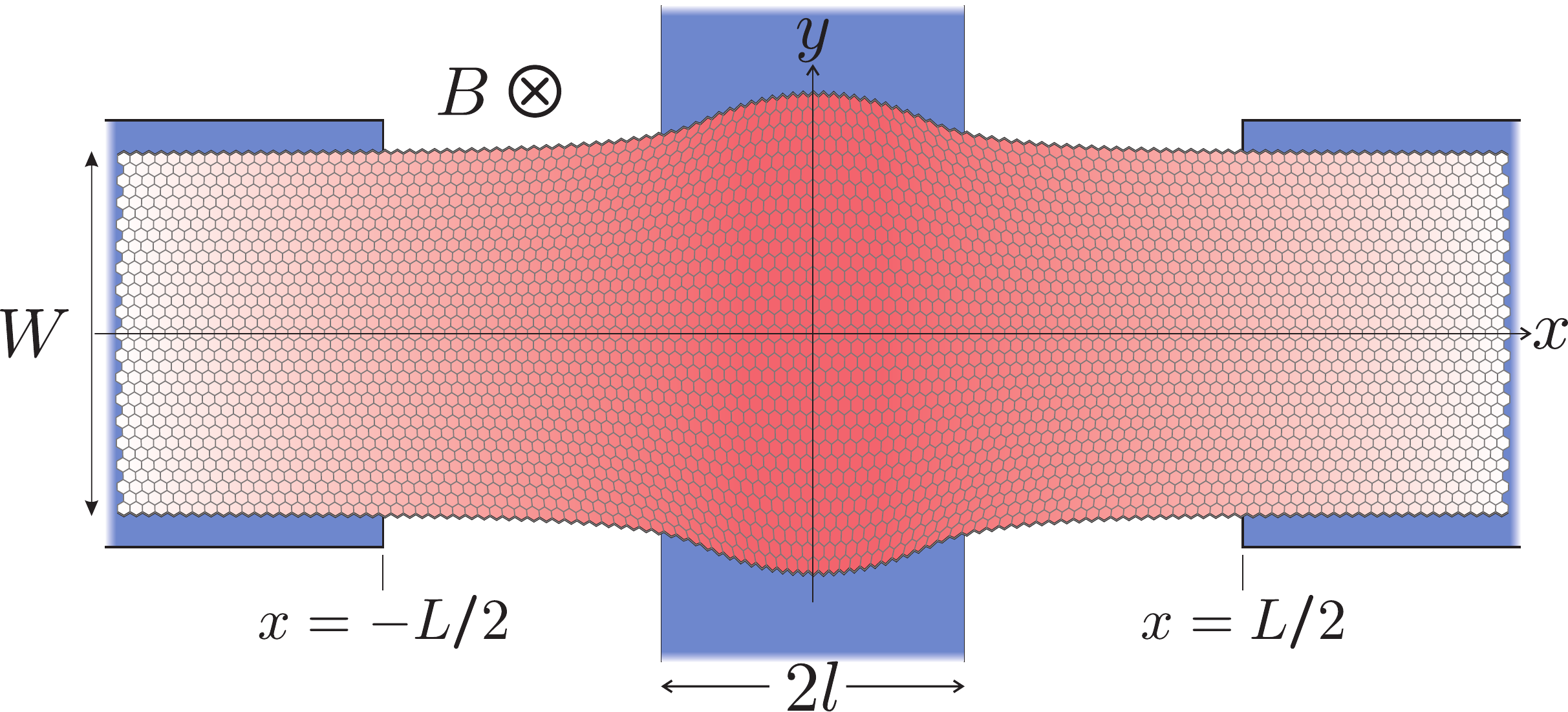}
\caption[fig]{Schematics of the system. The rectangles at the extremes of the $x$ and $y$ axes are the contacts, which are attached to a graphene sheet (the central object). The deformation of the sheet is strongly exaggerated.
The variable shading and honeycomb lattice are added for aesthetic purposes.
}
\label{fig:sketch}
\end{figure}

We model a suspended graphene sample as an elastic membrane occupying the rectangle $|x| < L / 2$, $|y| < W / 2$. The membrane has the two-dimensional (2D) Young modulus $C = 340\, {\text{N}}/{\text{m}} \approx 2100\, {\text{eV}}/{\text{nm}^2}$, the Poisson ratio $\sigma = 0.15$, and the shear modulus $\mu = C / [2 (1 + \sigma)] \approx 900\, {\text{eV}}/{\text{nm}^2}$. This membrane is supported by four contacts that have the same uniform height. We assume that the sample is in the state of the plain stress, i.e., we ignore the possibility of spontaneous wrinkling~\cite{CM03}. We model the two side contacts as rigid stamps of width $2 l < L$ centered at the symmetry axis $x = 0$, see Fig.~\ref{fig:sketch}. The remaining parts of the $y = \pm W / 2$ boundaries are assumed to be free of tractions. Each of the stamps pulls graphene normally outward with the total force $P$. Initially we will assume the stamps are rigidly clamped to the sheet. Later, we will
consider other boundary conditions. At the $x = \pm L / 2$ sides (the source and drain contacts) the sheet is clamped: $u = v = 0$. Here $u(x, y)$ and $v(x, y)$ are the elastic deformations in the $x$ and $y$-directions, respectively. At $P = 0$ the membrane is supposed to be flat and unstressed. We are interested in the deformation that develops if the force $P$ is finite.

Our program is as follows. First, we obtain an approximate solution of the posed elasticity theory problem. Next, we determine the distribution of the effective vector potential according to the formulas~\cite{M07, GHL08, Comment_on_valley}
\begin{equation}
\mathbb{A}_x = \beta\frac{\varphi}{a}(\partial_x u - \partial_y v)\,,
\quad
\mathbb{A}_y = -\beta\frac{\varphi}{a}(\partial_y u + \partial_x v)\,,
\label{eqn:A_Cartesian}
\end{equation}
where $\varphi = \hbar c / e$ is the flux quantum (reduced by $2 \pi$), $a = 0.14\,\text{nm}$ is the separation of the nearest carbon atoms and $\beta = d \ln t / d \ln a$ is the logarithmic derivative of the corresponding hopping integral $t \approx 3\,\text{eV}$.

The two Eqs. ~\eqref{eqn:A_Cartesian} can be combined using the complex-variable notation:
\begin{align}
\mathbb{A} &= \mathbb{A}_x + i \mathbb{A}_y = ({2 \beta \varphi} / {a}) \partial_z \overline{U}\,,
\quad U = u + i v\,,
\label{eqn:A}\\
\mathbb{B} &= \partial_x \mathbb{A}_y - \partial_y \mathbb{A}_x
 = 2\, \text{Im}\, \partial_z \mathbb{A}\,,
\quad z = x + i y\,,
\label{eqn:B_eff}
\end{align}
where the bar denotes complex conjugation. The last equation gives the pseudomagnetic field. As a final step, we calculate numerically the ballistic conductance between the source and drain contacts assuming the system is subject to the total magnetic field $B_\text{tot} = B + \mathbb{B}$.

%
% FIG. 2
%

Our elasticity theory problem does not seem to have an analytic solution but we can construct an approximate one from those for a half-plane. Consider therefore a membrane that occupies the entire lower half-plane $y < 0$. Let its boundary be unloaded except for the interval $-l < x < l$ in contact with a rigid stamp. The following general representation~\cite{Muskhelishvili1977sbp} is valid:
\begin{equation}
2 \mu U(z) = \varkappa \phi(z) + \phi(\bar{z}) + (\bar{z} - z)
\overline{\phi^\prime(z)}\,,
\quad \varkappa = \frac{3 - \sigma}{1 + \sigma}\,,
\label{eqn:U_from_phi}
\end{equation}
where function $\phi(z)$ and its derivative $\Phi(z) = \phi^\prime(z)$ are regular in the complex plane of $z$ with a branch cut $(-l, l)$. These functions are determined by the shape of the stamp and the forces it exerts on the membrane. For a stamp whose base is straight, parallel to the $x$-axis, and bonded to the membrane we have~\cite{Muskhelishvili1977sbp}
\begin{equation}
\Phi(z) = \frac{P}{2 \pi i}\, (z + l)^{-\frac12 + i m}
                               (z - l)^{-\frac12 - i m}
\label{eqn:Phi_straight}
\end{equation}
with $m = (\ln \varkappa) / (2 \pi) \approx 0.14$. The asymptotic behavior of $\Phi(z)$ at large $z = r e^{i \theta}$ is given by
\begin{equation}
\Phi(z) \simeq -2 i c_0 \frac{\mu}{\beta}\, \frac{a}{r} e^{-i \theta}\,,
\quad r \gg l\,,
\label{eqn:Phi_far}
\end{equation}
same as in the Flamant problem: a point force applied to a 2D elastic sheet~\cite{LL59}. Here we introduced the dimensionless coefficient $c_0 = \beta P / (4 \pi \mu a)$, for convenience. Substituting
Eq.~\eqref{eqn:Phi_far} into Eq.~\eqref{eqn:U_from_phi} and then to Eq.~\eqref{eqn:B_eff}, we find that $\mathbb{B}$ decays as the square of the distance and has a peculiar angular dependence:
\begin{equation}
\mathbb{B}(r, \theta) = \frac{c_0 \varphi}{r^2}
                \left({8 \cos 4 \theta -  4 \cos 2 \theta}\right)\,,
                \quad r \gg l\,.
\label{eqn:B_far}
\end{equation}
On the other hand, near the corners of the stamp, $z = \pm l$, the solution~\eqref{eqn:Phi_straight} is characterized by divergent oscillations. As discussed in Ref.~\cite{Muskhelishvili1977sbp}, such oscillations are unphysical. The linear elasticity theory usually fails at distances where they supposedly occur. Hence we consider the following alternative to
Eq.~\eqref{eqn:Phi_straight}:
\begin{equation}
\Phi(z) = \frac{P}{\pi i}\, \frac{1}{z + \sqrt{z^2 - l^2}}
 = \frac{P e^{-\zeta}}{\pi i l},
\quad
\zeta \equiv \arccos \frac{z}{l}\,.
\label{eqn:Phi_rounded}
\end{equation}
The corresponding elastic deformation is given by %%
\begin{equation}
U_1(z) = c_0 \frac{a}{\beta} \left(\varkappa \zeta - \bar{\zeta} +
\frac{\varkappa z}{i e^\zeta} + \frac{2 z - \bar{z}}{i e^{\bar{\zeta}}}
\right) + \text{const}\,.
\label{eqn:U_1}
\end{equation}
This happens to be the exact solution for a frictionless stamp with a rounded base~\cite{Muskhelishvili1977sbp}. It is physically relevant, free of divergences, and has the same universal far-field behavior, Eq~\eqref{eqn:Phi_far}. Since it is the far-field behavior that is important for the backscattering of the edge states in wide samples, we adopt Eq.~\eqref{eqn:U_1} as our basic building block. (The subscript ``1'' in $U_1$ is to remind us that it is for a single stamp.) We now construct the solution for the original problem simply as the sum~\cite{Comment_on_error}
\begin{equation}
U(z) = U_1(z - i W / 2) - U_1(-z - i W / 2)\,. 
\label{eqn:U}
\end{equation}
%%

%
% FIG. 3
%
\begin{figure}
\includegraphics[width=3.2in]{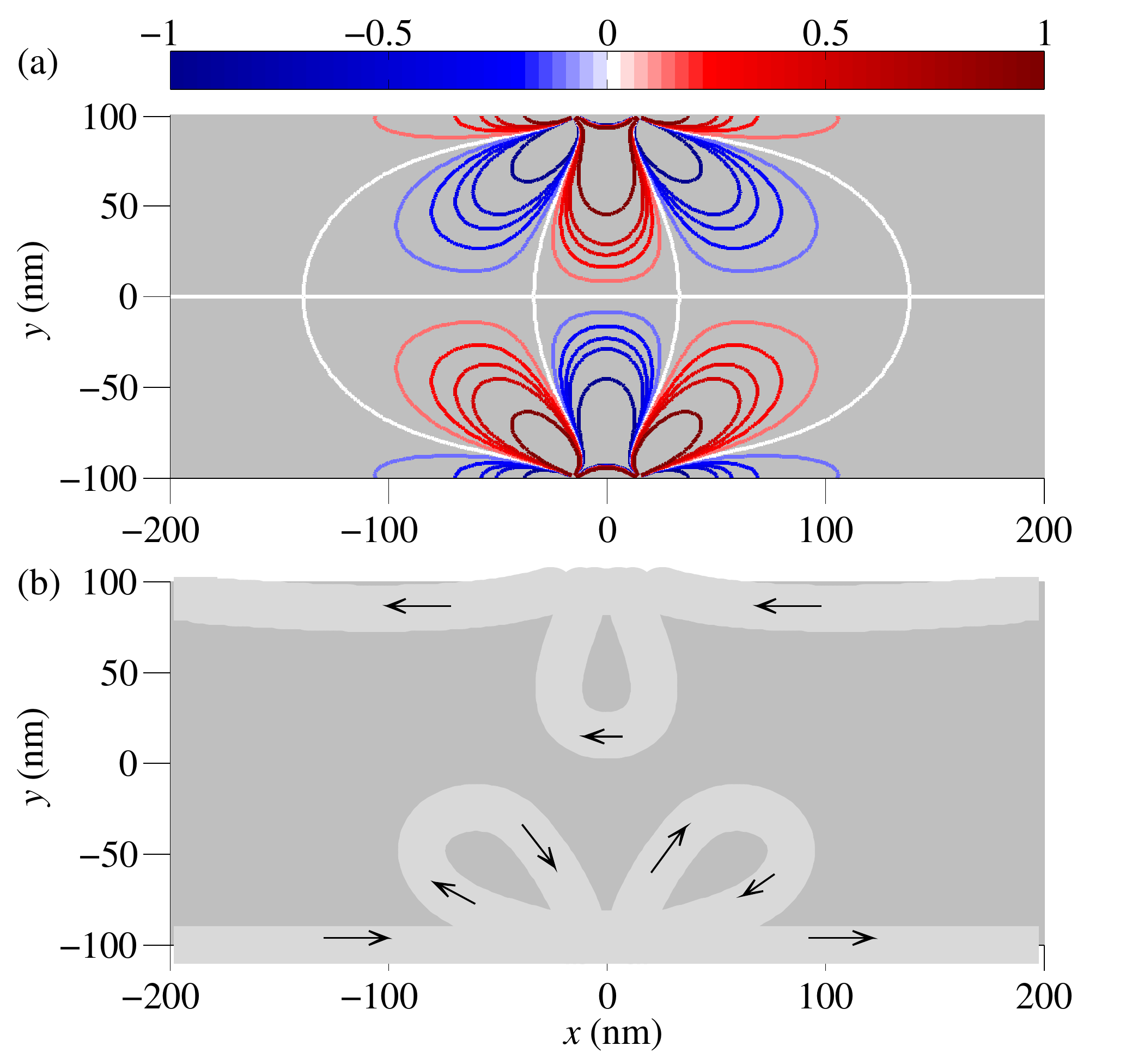}
\caption[fig]{(Color online). (a) Pseudomagnetic field, in Tesla,
induced by two stamps with parameters $2 l = 30\, \text{nm}$, $c_0 = 1$ in a sample of width $200\, \text{nm}$.
(b) Effect of this pseudomagnetic field on the edge states of the $N = 2$ Landau level at $B = 1\,\text{T}$. The edge states are depicted as light gray ribbons of thickness $\ell \approx 26\,\text{nm}$. The arrows indicate their propagation direction.}
\label{fig:B_eff}
\end{figure}

Let us now estimate the coefficient $c_0$ in Eq.~\eqref{eqn:U_1}. Up to logarithmic factors, $c_0 \sim \beta e_{y y} W / a$, where $e_{y y} = |U(i W) - U(0)| / W$ is the average strain. We see that the estimate~\cite{FGK08} of $e_{y y} \sim 10^{-4}$ for a sample of size $W \sim 1\,\mu\text{m}$ yields $c_0 \sim 1$. In this paper we are interested in samples of smaller size, $W \sim 100\, \text{nm}$. However, we think that $c_0 = 1$ is still a reasonable number if the pulling force $P$ is determined by the direct interaction of graphene with the contacts, not via the intermediary of the gate underneath. The distribution of the pseudomagnetic field computed for such $c_0$ and $W$ using Eqs.~\eqref{eqn:A}, \eqref{eqn:B_eff}, \eqref{eqn:U_1}, and \eqref{eqn:U} is shown in Fig.~\ref{fig:B_eff}. As one can see, $\mathbb{B} \sim 1\, \text{T}$. This suggests that strain can strongly affect magnetotransport if $B$ is comparable or smaller than $1\,\text{T}$. A more precise criterion can be derived as follows.

The suppression of the $N$th quantized plateau (centered at filling factor $4 N + 2$) is due to backscattering between the counter-propagating edge states of the $N$th Landau level. For $N \gg 1$ a semiclassical description is valid, in which the edge states are visualized as channels of width $\sim \ell \equiv \sqrt{\varphi / B}$ that follow the contours $\varepsilon_N = \hbar v_F \sqrt{{2 N} \left| B_\text{tot}(x, y) \right| / {\varphi} } = E_F$, where $v_F$ is the Fermi velocity and $E_F = \hbar v_F k_F$ is the Fermi energy. In the middle of the $N$th quantized plateau $E_F = \hbar v_F \sqrt{(2 N + 1) B / {\varphi}}\,$, so that the edge states follow the contours
\begin{equation}
\mathbb{B}(x, y) = {B} / (2 N) = \text{const}\,.
\label{eqn:edge_states}
\end{equation}
These contours eventually terminate at the sample boundaries because $\mathbb{B}$ decreases away from the stamps. Upon reaching these termination points, the edge states continue \textit{along\/} the boundaries, as usual [Fig.~\ref{fig:B_eff}(b)]. 

On the other hand, near the stamps the edge states veer into the bulk. The counter-propagating edge states can approach each other near the saddle points of $\mathbb{B}(x, y)$. Using Eq.~\eqref{eqn:B_far} one can show that for $2l \ll W$ the saddle-points closest to the origin are located at $y = 0$ and $x = \pm x_0$, where $x_0 = W (1 - 2 / \sqrt{5})^{1/2} / 2 \approx 0.16 W$. Near $x = x_0$ we find
\begin{equation}
\mathbb{B}(x, y) \simeq -c_0 c_1 \varphi (x - x_0) y / W^4\,,
\quad c_1
% = 300 \sqrt{50 + 22 \sqrt{5}}
\approx 2988\,.
\label{eqn:B_saddle}
\end{equation}
As the real magnetic field decreases, strong backscattering and therefore the complete destruction of the QHE is possible if the distance between the counter-propagating edge states at the saddle-point becomes $\sim \ell$. Using Eqs.~\eqref{eqn:edge_states} and \eqref{eqn:B_saddle}, this condition can be written as $2 W^2 / \ell^2 \sim \sqrt{c_0 c_1 N}$. Substituting $\ell = \sqrt{2 N + 1} / k_F$, we recover Eq.~\eqref{eqn:N_c}.

Now we discuss our numerical results for the ballistic conductance $G$ shown in
Fig.~\ref{fig:transmission}. The calculation was done by means of the recursive Green's function technique~\cite{FGA99}. The system was treated as a tight-binding model on a honeycomb lattice with nearest-neighbor hopping $t = t_0 \exp (-\beta |\delta a| / a)$. Here $\beta = 3.37$ (~\onlinecite{PNP09}) and $\delta a(z) = U(z + \tau) - U(z)$ is the complex-valued distortion of the bond that connects sites $z$ and $z + \tau$ in the unstrained lattice. Although we have referred to the side stamps as the contacts, in our calculations they are electrically isolated from the system, i.e., $G$ is the conductance between the source and the drain. Still, our results are representative of the four-contact conductance because we made sure that contact resistances of the source and the drain are small. In the $L \to \infty$ limit where the source and drain are infinitely far from the strained region these resistances vanish. We approached this limit by increasing $L$ while keeping $l$ and $W$ constant until no significant $L$-dependence of $G$ was seen~\cite{Comment_on_SD}. As one can see from Fig.~\ref{fig:transmission}, the calculations show the expected erosion of the QHE plateaus as $\nu$ increases and confirm the analytical estimate of the threshold filling factor $\nu_c$, Eq.~\eqref{eqn:N_c}.

%%%%%%%%%%%%%%%%%%%%%%%%%%%%%%%%%%%%%%%%%%%%%%%%%%%%%%%%%%%%%%%%%%%%
%{\em Discussion.}

In conclusion, we want to mention a few caveats:

(i) besides the pseudo vector potential $\mathbb{A}$, the strain also creates a scalar one~\cite{SA02b}. However due to screening, it is likely to be less important~\cite{FGK08, OGM09}.

(ii) We have assumed that electron transport is coherent. If dissipation is present, then the two-terminal conductance is better understood in terms of mixing of the Hall and longitudinal conductivities~\cite{Skachko2009iaf, FS94, AL09}. This effect is strongly enhanced in long samples~\cite{FS94}.

(iii) The point where a contact ends and a free edge begins can be considered a corner of angle $\alpha = \pi$ at which the boundary conditions suddenly change from clamped to free ones. Divergence of strain near such corners is well known. The exponent $\lambda$ of the corresponding power law depends on the Poisson ratio $\sigma$ and the corner angle $\alpha$. For $\alpha = \pi$ and $\sigma = 0.15$, Eq.~\eqref{eqn:Phi_straight} implies $\lambda = 1/2 \pm 0.14 i$. If we had instead $\alpha = \pi / 2$, it can be shown~\cite{Muskhelishvili1977sbp, Williams1952ssr} that $\lambda$ would be $0.16$. In this case, the strain at a distance of order $\ell$ from the corner is enhanced with respect to its value near the center of the flake by a factor of $(W / \ell)^\lambda$. For $\ell \sim 10\,\text{nm}$ and $W \sim 1\,\mu\text{m}$ this factor is $\sim 1.5$, which is a sizable effect. Accordingly, the recently proposed~\cite{PNP09, PN09, GKG09, FR09} ``strain engineering'' of graphene electronic devices must either take into account or take advantage of the strain singularities we discuss here.

%%%%%%%%%%%%%%%%%%%%%%%%%%%%%%%%%%%%%%%%%%%%%%%%%%%%%%%%%%%%%%%%%%%
GL and FG acknowledge support from MEC
(Spain) under grants FIS2008-00124 and CONSOLIDER CSD2007-00010,
and by the Comunidad de Madrid, under CITECNOMIK. EP and PS
acknowledge support from the European Commission, Marie Curie
Excellence Grant MEXT-CT-2005-023778. MF is supported by the NSF under Grant No. DMR-0706654.

\bibliography{side_strain_QHE}

%Merlin.mbs v4.21 2009-07-09.
\begin{thebibliography}{10}%
\makeatletter
\providecommand \@ifxundefined [1]{%
 \ifx #1\undefined \expandafter \@firstoftwo
 \else \expandafter \@secondoftwo
\fi
}%
\providecommand \@ifnum [1]{%
 \ifnum #1\expandafter \@firstoftwo
 \else \expandafter \@secondoftwo
\fi
}%
\providecommand \enquote [1]{``#1''}%
\providecommand \bibnamefont  [1]{#1}%
\providecommand \bibfnamefont [1]{#1}%
\providecommand \citenamefont [1]{#1}%
\providecommand\href[0]{\@sanitize\@href}%
\providecommand\@href[1]{\endgroup\@@startlink{#1}\endgroup\@@href}%
\providecommand\@@href[1]{#1\@@endlink}%
\providecommand \@sanitize [0]{\begingroup\catcode`\&12\catcode`\#12\relax}%
\@ifxundefined \pdfoutput {\@firstoftwo}{%
 \@ifnum{\z@=\pdfoutput}{\@firstoftwo}{\@secondoftwo}%
}{%
 \providecommand\@@startlink[1]{\leavevmode\special{html:<a href="#1">}}%
 \providecommand\@@endlink[0]{\special{html:</a>}}%
}{%
 \providecommand\@@startlink[1]{%
  \leavevmode
  \pdfstartlink
   attr{/Border[0 0 1 ]/H/I/C[0 1 1]}%
   user{/Subtype/Link/A<</Type/Action/S/URI/URI(#1)>>}%
  \relax
 }%
 \providecommand\@@endlink[0]{\pdfendlink}%
}%
\providecommand \url  [0]{\begingroup\@sanitize \@url }%
\providecommand \@url [1]{\endgroup\@href {#1}{\urlprefix}}%
\providecommand \urlprefix [0]{URL }%
\providecommand \Eprint[0]{\href }%
\@ifxundefined \urlstyle {%
  \providecommand \doi [1]{doi:\discretionary{}{}{}#1}%
}{%
  \providecommand \doi [0]{doi:\discretionary{}{}{}\begingroup
  \urlstyle{rm}\Url }%
}%
\providecommand \doibase [0]{http://dx.doi.org/}%
\providecommand \Doi[1]{\href{\doibase#1}}%
\providecommand \bibAnnote [3]{%
  \BibitemShut{#1}%
  \begin{quotation}\noindent
    \textsc{Key:}\ #2\\\textsc{Annotation:}\ #3%
  \end{quotation}%
}%
\providecommand \bibAnnoteFile [2]{%
  \IfFileExists{#2}{\bibAnnote {#1} {#2} {\input{#2}}}{}%
}%
\providecommand \typeout [0]{\immediate \write \m@ne }%
\providecommand \selectlanguage [0]{\@gobble}%
\providecommand \bibinfo [0]{\@secondoftwo}%
\providecommand \bibfield [0]{\@secondoftwo}%
\providecommand \translation [1]{[#1]}%
\providecommand \BibitemOpen[0]{}%
\providecommand \bibitemStop [0]{}%
\providecommand \bibitemNoStop [0]{.\EOS\space}%
\providecommand \EOS [0]{\spacefactor3000\relax}%
\providecommand \BibitemShut [1]{\csname bibitem#1\endcsname}%
%</preamble>
\bibitem{Netal04}%
  \BibitemOpen
  \bibfield{author}{%
  \bibinfo {author} {\bibfnamefont{K.~S.}\ \bibnamefont{Novoselov}}, \bibinfo
  {author} {\bibfnamefont{A.~K.}\ \bibnamefont{Geim}}, \bibinfo {author}
  {\bibfnamefont{S.~V.}\ \bibnamefont{Morozov}}, \bibinfo {author}
  {\bibfnamefont{D.}~\bibnamefont{Jiang}}, \bibinfo {author}
  {\bibfnamefont{Y.}~\bibnamefont{Zhang}}, \bibinfo {author}
  {\bibfnamefont{S.~V.}\ \bibnamefont{Dubonos}}, \bibinfo {author}
  {\bibfnamefont{I.~V.}\ \bibnamefont{Grigorieva}},\ and\ \bibinfo {author}
  {\bibfnamefont{A.~A.}\ \bibnamefont{Firsov}},\ }%
  \bibfield{journal}{%
  \bibinfo {journal} {Science}\ }%
  \textbf{\bibinfo {volume} {306}},\ \bibinfo {pages} {666} (\bibinfo {year}
  {2004})%
  \bibAnnoteFile{NoStop}{Netal04}%
\bibitem{Netal05}%
  \BibitemOpen
  \bibfield{author}{%
  \bibinfo {author} {\bibfnamefont{K.~S.}\ \bibnamefont{Novoselov}}, \bibinfo
  {author} {\bibfnamefont{D.}~\bibnamefont{Jiang}}, \bibinfo {author}
  {\bibfnamefont{F.}~\bibnamefont{Schedin}}, \bibinfo {author}
  {\bibfnamefont{T.~J.}\ \bibnamefont{Booth}}, \bibinfo {author}
  {\bibfnamefont{V.~V.}\ \bibnamefont{Khotkevich}}, \bibinfo {author}
  {\bibfnamefont{S.~V.}\ \bibnamefont{Morozov}},\ and\ \bibinfo {author}
  {\bibfnamefont{A.~K.}\ \bibnamefont{Geim}},\ }%
  \bibfield{journal}{%
  \bibinfo {journal} {Proc. Natl. Acad. Sci. U.S.A.}\ }%
  \textbf{\bibinfo {volume} {102}},\ \bibinfo {pages} {10451} (\bibinfo {year}
  {2005})%
  \bibAnnoteFile{NoStop}{Netal05}%
\bibitem{Netal05b}%
  \BibitemOpen
  \bibfield{author}{%
  \bibinfo {author} {\bibfnamefont{K.~S.}\ \bibnamefont{Novoselov}}, \bibinfo
  {author} {\bibfnamefont{A.~K.}\ \bibnamefont{Geim}}, \bibinfo {author}
  {\bibfnamefont{S.~V.}\ \bibnamefont{Morozov}}, \bibinfo {author}
  {\bibfnamefont{D.}~\bibnamefont{Jiang}}, \bibinfo {author}
  {\bibfnamefont{M.~I.}\ \bibnamefont{Katsnelson}}, \bibinfo {author}
  {\bibfnamefont{I.~V.}\ \bibnamefont{Grigorieva}}, \bibinfo {author}
  {\bibfnamefont{S.~V.}\ \bibnamefont{Dubonos}},\ and\ \bibinfo {author}
  {\bibfnamefont{A.~A.}\ \bibnamefont{Firsov}},\ }%
  \bibfield{journal}{%
  \bibinfo {journal} {Nature}\ }%
  \textbf{\bibinfo {volume} {438}},\ \bibinfo {pages} {197} (\bibinfo {year}
  {2005})%
  \bibAnnoteFile{NoStop}{Netal05b}%
\bibitem{ZTSK05}%
  \BibitemOpen
  \bibfield{author}{%
  \bibinfo {author} {\bibfnamefont{Y.}~\bibnamefont{Zhang}}, \bibinfo {author}
  {\bibfnamefont{Y.-W.}\ \bibnamefont{Tan}}, \bibinfo {author}
  {\bibfnamefont{H.~L.}\ \bibnamefont{Stormer}},\ and\ \bibinfo {author}
  {\bibfnamefont{P.}~\bibnamefont{Kim}},\ }%
  \bibfield{journal}{%
  \bibinfo {journal} {Nature}\ }%
  \textbf{\bibinfo {volume} {438}},\ \bibinfo {pages} {201} (\bibinfo {year}
  {2005})%
  \bibAnnoteFile{NoStop}{ZTSK05}%
\bibitem{NGPNG09}%
  \BibitemOpen
  \bibfield{author}{%
  \bibinfo {author} {\bibfnamefont{A.~H.}\ \bibnamefont{{Castro Neto}}},
  \bibinfo {author} {\bibfnamefont{F.}~\bibnamefont{Guinea}}, \bibinfo {author}
  {\bibfnamefont{N.~M.~R.}\ \bibnamefont{Peres}}, \bibinfo {author}
  {\bibfnamefont{K.~S.}\ \bibnamefont{Novoselov}},\ and\ \bibinfo {author}
  {\bibfnamefont{A.~K.}\ \bibnamefont{Geim}},\ }%
  \bibfield{journal}{%
  \bibinfo {journal} {Rev. Mod. Phys.}\ }%
  \textbf{\bibinfo {volume} {81}},\ \bibinfo {pages} {109} (\bibinfo {year}
  {2009})%
  \bibAnnoteFile{NoStop}{NGPNG09}%
\bibitem{AGSci09}%
  \BibitemOpen
  \bibfield{author}{%
  \bibinfo {author} {\bibfnamefont{A.~K.}\ \bibnamefont{Geim}},\ }%
  \bibfield{journal}{%
  \bibinfo {journal} {Science}\ }%
  \textbf{\bibinfo {volume} {324}},\ \bibinfo {pages} {1530} (\bibinfo {year}
  {2009})%
  \bibAnnoteFile{NoStop}{AGSci09}%
\bibitem{Netal07}%
  \BibitemOpen
  \bibfield{author}{%
  \bibinfo {author} {\bibfnamefont{K.~S.}\ \bibnamefont{Novoselov}}, \bibinfo
  {author} {\bibfnamefont{Z.}~\bibnamefont{Jiang}}, \bibinfo {author}
  {\bibfnamefont{Y.}~\bibnamefont{Zhang}}, \bibinfo {author}
  {\bibfnamefont{S.~V.}\ \bibnamefont{Morozov}}, \bibinfo {author}
  {\bibfnamefont{H.~L.}\ \bibnamefont{Stormer}}, \bibinfo {author}
  {\bibfnamefont{U.}~\bibnamefont{Zeitler}}, \bibinfo {author}
  {\bibfnamefont{J.~C.}\ \bibnamefont{Maan}}, \bibinfo {author}
  {\bibfnamefont{G.~S.}\ \bibnamefont{Boebinger}}, \bibinfo {author}
  {\bibfnamefont{P.}~\bibnamefont{Kim}},\ and\ \bibinfo {author}
  {\bibfnamefont{A.~K.}\ \bibnamefont{Geim}},\ }%
  \bibfield{journal}{%
  \bibinfo {journal} {Science}\ }%
  \textbf{\bibinfo {volume} {315}},\ \bibinfo {pages} {1379} (\bibinfo {year}
  {2007})%
  \bibAnnoteFile{NoStop}{Netal07}%
\bibitem{Betal08a}%
  \BibitemOpen
  \bibfield{author}{%
  \bibinfo {author} {\bibfnamefont{K.~I.}\ \bibnamefont{Bolotin}}, \bibinfo
  {author} {\bibfnamefont{K.~J.}\ \bibnamefont{Sikes}}, \bibinfo {author}
  {\bibfnamefont{Z.}~\bibnamefont{Jiang}}, \bibinfo {author}
  {\bibfnamefont{G.}~\bibnamefont{Fudenberg}}, \bibinfo {author}
  {\bibfnamefont{J.}~\bibnamefont{Hone}}, \bibinfo {author}
  {\bibfnamefont{P.}~\bibnamefont{Kim}},\ and\ \bibinfo {author}
  {\bibfnamefont{H.~L.}\ \bibnamefont{Stormer}},\ }%
  \bibfield{journal}{%
  \bibinfo {journal} {Solid State Commun.}\ }%
  \textbf{\bibinfo {volume} {156}},\ \bibinfo {pages} {351} (\bibinfo {year}
  {2008})%
  \bibAnnoteFile{NoStop}{Betal08a}%
\bibitem{DSBA08}%
  \BibitemOpen
  \bibfield{author}{%
  \bibinfo {author} {\bibfnamefont{X.}~\bibnamefont{Du}}, \bibinfo {author}
  {\bibfnamefont{I.}~\bibnamefont{Skachko}}, \bibinfo {author}
  {\bibfnamefont{A.}~\bibnamefont{Barker}},\ and\ \bibinfo {author}
  {\bibfnamefont{E.~Y.}\ \bibnamefont{Andrei}},\ }%
  \bibfield{journal}{%
  \bibinfo {journal} {Nat. Nanotech.}\ }%
  \textbf{\bibinfo {volume} {3}},\ \bibinfo {pages} {491} (\bibinfo {year}
  {2008})%
  \bibAnnoteFile{NoStop}{DSBA08}%
\bibitem{BSHSK08}%
  \BibitemOpen
  \bibfield{author}{%
  \bibinfo {author} {\bibfnamefont{K.~I.}\ \bibnamefont{Bolotin}}, \bibinfo
  {author} {\bibfnamefont{K.~J.}\ \bibnamefont{Sikes}}, \bibinfo {author}
  {\bibfnamefont{J.}~\bibnamefont{Hone}}, \bibinfo {author}
  {\bibfnamefont{H.~L.}\ \bibnamefont{Stormer}},\ and\ \bibinfo {author}
  {\bibfnamefont{P.}~\bibnamefont{Kim}},\ }%
  \bibfield{journal}{%
  \bibinfo {journal} {Phys. Rev. Lett.}\ }%
  \textbf{\bibinfo {volume} {101}},\ \bibinfo {pages} {096802} (\bibinfo {year}
  {2008})%
  \bibAnnoteFile{NoStop}{BSHSK08}%
\bibitem{Skachko2009iaf}%
  \BibitemOpen
  \bibfield{author}{%
  \bibinfo {author} {\bibfnamefont{I.}~\bibnamefont{Skachko}}, \bibinfo
  {author} {\bibfnamefont{X.}~\bibnamefont{Du}}, \bibinfo {author}
  {\bibfnamefont{F.}~\bibnamefont{Duerr}}, \bibinfo {author}
  {\bibfnamefont{A.}~\bibnamefont{Luican}}, \bibinfo {author}
  {\bibfnamefont{D.~A.}\ \bibnamefont{Abanin}}, \bibinfo {author}
  {\bibfnamefont{L.~S.}\ \bibnamefont{Levitov}},\ and\ \bibinfo {author}
  {\bibfnamefont{E.~Y.}\ \bibnamefont{Andrei}},\ }%
  \bibinfo {note} {arXiv:0910.2518 (unpublished)}%
  \bibAnnoteFile{NoStop}{Skachko2009iaf}%
\bibitem{Bolotin2009oot}%
  \BibitemOpen
  \bibfield{author}{%
  \bibinfo {author} {\bibfnamefont{K.~I.}\ \bibnamefont{Bolotin}}, \bibinfo
  {author} {\bibfnamefont{F.}~\bibnamefont{Ghahari}}, \bibinfo {author}
  {\bibfnamefont{M.~D.}\ \bibnamefont{Shulman}}, \bibinfo {author}
  {\bibfnamefont{H.~L.}\ \bibnamefont{Stormer}},\ and\ \bibinfo {author}
  {\bibfnamefont{P.}~\bibnamefont{Kim}},\ }%
  \bibfield{journal}{%
  \Doi{10.1038/nature08582}{\bibinfo {journal} {Nature}}\ }%
  \textbf{\bibinfo {volume} {462}},\ \bibinfo {pages} {196} (\bibinfo {month}
  {Nov.}\ \bibinfo {year} {2009})%
  \bibAnnoteFile{NoStop}{Bolotin2009oot}%
\bibitem{FMY09}%
  \BibitemOpen
  \bibfield{author}{%
  \bibinfo {author} {\bibfnamefont{B.~E.}\ \bibnamefont{Feldman}}, \bibinfo
  {author} {\bibfnamefont{J.}~\bibnamefont{Martin}},\ and\ \bibinfo {author}
  {\bibfnamefont{A.}~\bibnamefont{Yacoby}},\ }%
  \bibfield{journal}{%
  \bibinfo {journal} {Nat. Phys.}\ }%
  \textbf{\bibinfo {volume} {5}},\ \bibinfo {pages} {889} (\bibinfo {year}
  {2009})%
  \bibAnnoteFile{NoStop}{FMY09}%
\bibitem{McEuen1990nrf}%
  \BibitemOpen
  \bibfield{author}{%
  \bibinfo {author} {\bibfnamefont{P.~L.}\ \bibnamefont{McEuen}}, \bibinfo
  {author} {\bibfnamefont{A.}~\bibnamefont{Szafer}}, \bibinfo {author}
  {\bibfnamefont{C.~A.}\ \bibnamefont{Richter}}, \bibinfo {author}
  {\bibfnamefont{B.~W.}\ \bibnamefont{Alphenaar}}, \bibinfo {author}
  {\bibfnamefont{J.~K.}\ \bibnamefont{Jain}}, \bibinfo {author}
  {\bibfnamefont{A.~D.}\ \bibnamefont{Stone}}, \bibinfo {author}
  {\bibfnamefont{R.~G.}\ \bibnamefont{Wheeler}},\ and\ \bibinfo {author}
  {\bibfnamefont{R.~N.}\ \bibnamefont{Sacks}},\ }%
  \bibfield{journal}{%
  \Doi{10.1103/PhysRevLett.64.2062}{\bibinfo {journal} {Phys. Rev. Lett.}}\ }%
  \textbf{\bibinfo {volume} {64}},\ \bibinfo {pages} {2062} (\bibinfo {month}
  {Apr}\ \bibinfo {year} {1990})%
  \bibAnnoteFile{NoStop}{McEuen1990nrf}%
\bibitem{Richter1992ooq}%
  \BibitemOpen
  \bibfield{author}{%
  \bibinfo {author} {\bibfnamefont{C.}~\bibnamefont{Richter}}, \bibinfo
  {author} {\bibfnamefont{R.}~\bibnamefont{Wheeler}},\ and\ \bibinfo {author}
  {\bibfnamefont{R.}~\bibnamefont{Sacks}},\ }%
  \bibfield{journal}{%
  \Doi{DOI: 10.1016/0039-6028(92)90350-F}{\bibinfo {journal} {Surf. Sci.}}\ }%
  \textbf{\bibinfo {volume} {263}},\ \bibinfo {pages} {270} (\bibinfo {year}
  {1992})%
  \bibAnnoteFile{NoStop}{Richter1992ooq}%
\bibitem{FS94}%
  \BibitemOpen
  \bibfield{author}{%
  \bibinfo {author} {\bibfnamefont{M.~M.}\ \bibnamefont{Fogler}}\ and\ \bibinfo
  {author} {\bibfnamefont{B.~I.}\ \bibnamefont{Shklovskii}},\ }%
  \bibfield{journal}{%
  \bibinfo {journal} {Phys. Rev. B}\ }%
  \textbf{\bibinfo {volume} {50}},\ \bibinfo {pages} {1656} (\bibinfo {year}
  {1994})%
  \bibAnnoteFile{NoStop}{FS94}%
\bibitem{Shlimak2004cow}%
  \BibitemOpen
  \bibfield{author}{%
  \bibinfo {author} {\bibfnamefont{I.}~\bibnamefont{Shlimak}}, \bibinfo
  {author} {\bibfnamefont{V.}~\bibnamefont{Ginodman}}, \bibinfo {author}
  {\bibfnamefont{M.}~\bibnamefont{Levin}}, \bibinfo {author}
  {\bibfnamefont{M.}~\bibnamefont{Potemski}}, \bibinfo {author}
  {\bibfnamefont{D.~K.}\ \bibnamefont{Maude}}, \bibinfo {author}
  {\bibfnamefont{A.}~\bibnamefont{Gerber}}, \bibinfo {author}
  {\bibfnamefont{A.}~\bibnamefont{Milner}},\ and\ \bibinfo {author}
  {\bibfnamefont{D.~J.}\ \bibnamefont{Paul}},\ }%
  \bibfield{journal}{%
  \Doi{10.1002/pssc.200303643}{\bibinfo {journal} {Phys. Status Sol. C}}\ }%
  \textbf{\bibinfo {volume} {1}},\ \bibinfo {pages} {67} (\bibinfo {year}
  {2004})%
  \bibAnnoteFile{NoStop}{Shlimak2004cow}%
\bibitem{SA02b}%
  \BibitemOpen
  \bibfield{author}{%
  \bibinfo {author} {\bibfnamefont{H.}~\bibnamefont{Suzuura}}\ and\ \bibinfo
  {author} {\bibfnamefont{T.}~\bibnamefont{Ando}},\ }%
  \bibfield{journal}{%
  \bibinfo {journal} {Phys. Rev. B}\ }%
  \textbf{\bibinfo {volume} {65}},\ \bibinfo {pages} {235412} (\bibinfo {year}
  {2002})%
  \bibAnnoteFile{NoStop}{SA02b}%
\bibitem{M07}%
  \BibitemOpen
  \bibfield{author}{%
  \bibinfo {author} {\bibfnamefont{J.~L.}\ \bibnamefont{{Ma\~nes}}},\ }%
  \bibfield{journal}{%
  \bibinfo {journal} {Phys. Rev. B}\ }%
  \textbf{\bibinfo {volume} {76}},\ \bibinfo {pages} {045430} (\bibinfo {year}
  {2007})%
  \bibAnnoteFile{NoStop}{M07}%
\bibitem{FGK08}%
  \BibitemOpen
  \bibfield{author}{%
  \bibinfo {author} {\bibfnamefont{M.~M.}\ \bibnamefont{Fogler}}, \bibinfo
  {author} {\bibfnamefont{F.}~\bibnamefont{Guinea}},\ and\ \bibinfo {author}
  {\bibfnamefont{M.~I.}\ \bibnamefont{Katsnelson}},\ }%
  \bibfield{journal}{%
  \bibinfo {journal} {Phys. Rev. Lett.}\ }%
  \textbf{\bibinfo {volume} {101}},\ \bibinfo {pages} {226804} (\bibinfo {year}
  {2008})%
  \bibAnnoteFile{NoStop}{FGK08}%
\bibitem{WPE09}%
  \BibitemOpen
  \bibfield{author}{%
  \bibinfo {author} {\bibfnamefont{Z.}~\bibnamefont{Wang}}, \bibinfo {author}
  {\bibfnamefont{L.}~\bibnamefont{Philippe}},\ and\ \bibinfo {author}
  {\bibfnamefont{J.}~\bibnamefont{Elias}},\ }%
  \bibinfo {note} {arXiv:0909.3428 (unpublished)}%
  \bibAnnoteFile{NoStop}{WPE09}%
\bibitem{Betal08b}%
  \BibitemOpen
  \bibfield{author}{%
  \bibinfo {author} {\bibfnamefont{J.~S.}\ \bibnamefont{Bunch}}, \bibinfo
  {author} {\bibfnamefont{S.~S.}\ \bibnamefont{Verbridge}}, \bibinfo {author}
  {\bibfnamefont{J.~S.}\ \bibnamefont{Alden}}, \bibinfo {author}
  {\bibfnamefont{A.~M.}\ \bibnamefont{{van der Zande}}}, \bibinfo {author}
  {\bibfnamefont{J.~M.}\ \bibnamefont{Parpia}}, \bibinfo {author}
  {\bibfnamefont{H.~G.}\ \bibnamefont{Craighead}},\ and\ \bibinfo {author}
  {\bibfnamefont{P.~L.}\ \bibnamefont{McEuen}},\ }%
  \bibfield{journal}{%
  \bibinfo {journal} {Nano Lett.}\ }%
  \textbf{\bibinfo {volume} {8}},\ \bibinfo {pages} {2458} (\bibinfo {year}
  {2008})%
  \bibAnnoteFile{NoStop}{Betal08b}%
\bibitem{Betal09}%
  \BibitemOpen
  \bibfield{author}{%
  \bibinfo {author} {\bibfnamefont{W.}~\bibnamefont{Bao}}, \bibinfo {author}
  {\bibfnamefont{F.}~\bibnamefont{Miao}}, \bibinfo {author}
  {\bibfnamefont{Z.}~\bibnamefont{Chen}}, \bibinfo {author}
  {\bibfnamefont{H.}~\bibnamefont{Zhang}}, \bibinfo {author}
  {\bibfnamefont{W.}~\bibnamefont{Jang}}, \bibinfo {author}
  {\bibfnamefont{C.}~\bibnamefont{Dames}},\ and\ \bibinfo {author}
  {\bibfnamefont{C.~N.}\ \bibnamefont{Lau}},\ }%
  \bibfield{journal}{%
  \bibinfo {journal} {Nat. Nanotech.}\ }%
  \textbf{\bibinfo {volume} {4}},\ \bibinfo {pages} {562} (\bibinfo {year}
  {2009})%
  \bibAnnoteFile{NoStop}{Betal09}%
\bibitem{Tetal09}%
  \BibitemOpen
  \bibfield{author}{%
  \bibinfo {author} {\bibfnamefont{M.~L.}\ \bibnamefont{Teague}}, \bibinfo
  {author} {\bibfnamefont{A.~P.}\ \bibnamefont{Lai}}, \bibinfo {author}
  {\bibfnamefont{J.}~\bibnamefont{Velasco}}, \bibinfo {author}
  {\bibfnamefont{C.~R.}\ \bibnamefont{Hughes}}, \bibinfo {author}
  {\bibfnamefont{A.~D.}\ \bibnamefont{Beyer}}, \bibinfo {author}
  {\bibfnamefont{M.~W.}\ \bibnamefont{Bockrath}}, \bibinfo {author}
  {\bibfnamefont{C.~N.}\ \bibnamefont{Lau}},\ and\ \bibinfo {author}
  {\bibfnamefont{N.-C.}\ \bibnamefont{Yeh}},\ }%
  \bibfield{journal}{%
  \bibinfo {journal} {Nano Lett.}\ }%
  \textbf{\bibinfo {volume} {9}},\ \bibinfo {pages} {2542} (\bibinfo {year}
  {2009})%
  \bibAnnoteFile{NoStop}{Tetal09}%
\bibitem{PNP09}%
  \BibitemOpen
  \bibfield{author}{%
  \bibinfo {author} {\bibfnamefont{V.~M.}\ \bibnamefont{Pereira}}, \bibinfo
  {author} {\bibfnamefont{A.~H.}\ \bibnamefont{{Castro Neto}}},\ and\ \bibinfo
  {author} {\bibfnamefont{N.~M.~R.}\ \bibnamefont{Peres}},\ }%
  \bibfield{journal}{%
  \bibinfo {journal} {Phys. Rev. B}\ }%
  \textbf{\bibinfo {volume} {80}},\ \bibinfo {pages} {045401} (\bibinfo {year}
  {2009})%
  \bibAnnoteFile{NoStop}{PNP09}%
\bibitem{PN09}%
  \BibitemOpen
  \bibfield{author}{%
  \bibinfo {author} {\bibfnamefont{V.~M.}\ \bibnamefont{Pereira}}\ and\
  \bibinfo {author} {\bibfnamefont{A.~H.}\ \bibnamefont{{Castro Neto}}},\ }%
  \bibfield{journal}{%
  \bibinfo {journal} {Phys. Rev. Lett.}\ }%
  \textbf{\bibinfo {volume} {103}},\ \bibinfo {pages} {046801} (\bibinfo {year}
  {2009})%
  \bibAnnoteFile{NoStop}{PN09}%
\bibitem{GKG09}%
  \BibitemOpen
  \bibfield{author}{%
  \bibinfo {author} {\bibfnamefont{F.}~\bibnamefont{Guinea}}, \bibinfo {author}
  {\bibfnamefont{M.~I.}\ \bibnamefont{Katsnelson}},\ and\ \bibinfo {author}
  {\bibfnamefont{A.~K.}\ \bibnamefont{Geim}},\ }%
  \bibfield{journal}{%
  \bibinfo {journal} {Nat. Phys.}\ }%
  \textbf{\bibinfo {volume} {6}},\ \bibinfo {pages} {30} (\bibinfo {year}
  {2009})%
  \bibAnnoteFile{NoStop}{GKG09}%
\bibitem{FR09}%
  \BibitemOpen
  \bibfield{author}{%
  \bibinfo {author} {\bibfnamefont{M.}~\bibnamefont{Farjam}}\ and\ \bibinfo
  {author} {\bibfnamefont{H.}~\bibnamefont{Rafii-Tabar}},\ }%
  \bibinfo {note} {arXiv:0909.5052 (unpublished)}%
  \bibAnnoteFile{NoStop}{FR09}%
\bibitem{CM03}%
  \BibitemOpen
  \bibfield{author}{%
  \bibinfo {author} {\bibfnamefont{E.}~\bibnamefont{Cerda}}\ and\ \bibinfo
  {author} {\bibfnamefont{L.}~\bibnamefont{Mahadevan}},\ }%
  \bibfield{journal}{%
  \bibinfo {journal} {Phys. Rev. Lett.}\ }%
  \textbf{\bibinfo {volume} {90}},\ \bibinfo {pages} {074302} (\bibinfo {year}
  {2003})%
  \bibAnnoteFile{NoStop}{CM03}%
\bibitem{GHL08}%
  \BibitemOpen
  \bibfield{author}{%
  \bibinfo {author} {\bibfnamefont{F.}~\bibnamefont{Guinea}}, \bibinfo {author}
  {\bibfnamefont{B.}~\bibnamefont{Horovitz}},\ and\ \bibinfo {author}
  {\bibfnamefont{P.~L.}\ \bibnamefont{Doussal}},\ }%
  \bibfield{journal}{%
  \bibinfo {journal} {Phys. Rev. B}\ }%
  \textbf{\bibinfo {volume} {77}},\ \bibinfo {pages} {205421} (\bibinfo {year}
  {2008})%
  \bibAnnoteFile{NoStop}{GHL08}%
\bibitem{Comment_on_valley}%
  \BibitemOpen
  \bibinfo {note} {Here we assume that the $x$-axis is along the zigzag
  direction. The formulas for $\mathbb{A}$ correspond to the valley $K = (-4
  \pi\, /\, 3 \sqrt{3} a, 0)$. The vector potential for the valley $-K$ has the
  opposite sign.}%
  \bibAnnoteFile{Stop}{Comment_on_valley}%
\bibitem{Muskhelishvili1977sbp}%
  \BibitemOpen
  \bibfield{author}{%
  \bibinfo {author} {\bibfnamefont{N.~I.}\ \bibnamefont{Muskhelishvili}},\ }%
  \emph{\bibinfo {title} {Some Basic Problems of the Mathematical Theory of
  Elasticity}}\ (\bibinfo {publisher} {Springer},\ \bibinfo {address} {New
  York},\ \bibinfo {year} {1977})%
  \bibAnnoteFile{NoStop}{Muskhelishvili1977sbp}%
\bibitem{LL59}%
  \BibitemOpen
  \bibfield{author}{%
  \bibinfo {author} {\bibfnamefont{L.~D.}\ \bibnamefont{Landau}}\ and\ \bibinfo
  {author} {\bibfnamefont{E.~M.}\ \bibnamefont{Lifschitz}},\ }%
  \emph{\bibinfo {title} {Theory of Elasticity}}\ (\bibinfo {publisher}
  {Pergamon Press, Oxford},\ \bibinfo {year} {1959})%
  \bibAnnoteFile{NoStop}{LL59}%
\bibitem{Comment_on_error}%
  \BibitemOpen
  \bibinfo {note} {The error made in this approximation is presumably no
  greater than the error involved in approximating the real shape of the sample
  by a simple rectangle.}%
  \bibAnnoteFile{Stop}{Comment_on_error}%
\bibitem{FGA99}%
  \BibitemOpen
  \bibfield{author}{%
  \bibinfo {author} {\bibfnamefont{D.}~\bibnamefont{Ferry}}, \bibinfo {author}
  {\bibfnamefont{S.}~\bibnamefont{Goodnick}},\ and\ \bibinfo {author}
  {\bibfnamefont{H.}~\bibnamefont{Ahmad}},\ }%
  \emph{\bibinfo {title} {Transport in Nanostructures}}\ (\bibinfo {publisher}
  {Cambridge University Press},\ \bibinfo {address} {Cambridge},\ \bibinfo
  {year} {1999})%
  \bibAnnoteFile{NoStop}{FGA99}%
\bibitem{Comment_on_SD}%
  \BibitemOpen
  \bibinfo {note} {In any case, the successful recent observation of the QHE in
  the two-terminal setup~\cite{DSBA08, Skachko2009iaf, Bolotin2009oot} suggest
  that the source (drain) contact resistance is not the issue. Theoretical
  study of transport in zero magnetic field~\cite{FGK08} point to the same
  conclusion: the pseudomagnetic field at the source (drain) contacts becomes a
  significant perturbation only at very small densities or very large strains,
  and is further relaxed by residual impurities.}%
  \bibAnnoteFile{Stop}{Comment_on_SD}%
\bibitem{OGM09}%
  \BibitemOpen
  \bibfield{author}{%
  \bibinfo {author} {\bibfnamefont{F.}~\bibnamefont{von Oppen}}, \bibinfo
  {author} {\bibfnamefont{F.}~\bibnamefont{Guinea}},\ and\ \bibinfo {author}
  {\bibfnamefont{E.}~\bibnamefont{Mariani}},\ }%
  \bibfield{journal}{%
  \bibinfo {journal} {Phys. Rev. B}\ }%
  \textbf{\bibinfo {volume} {80}},\ \bibinfo {pages} {075420} (\bibinfo {year}
  {2009})%
  \bibAnnoteFile{NoStop}{OGM09}%
\bibitem{AL09}%
  \BibitemOpen
  \bibfield{author}{%
  \bibinfo {author} {\bibfnamefont{D.}~\bibnamefont{Abanin}}, \bibinfo {author}
  {\bibfnamefont{A.}~\bibnamefont{Shytov}}, \bibinfo {author}
  {\bibfnamefont{L.}~\bibnamefont{Levitov}},\ and\ \bibinfo {author}
  {\bibfnamefont{B.}~\bibnamefont{Halperin}},\ }%
  \bibfield{journal}{%
  \bibinfo {journal} {Phys. Rev. B}\ }%
  \textbf{\bibinfo {volume} {79}},\ \bibinfo {pages} {035304} (\bibinfo {year}
  {2009})%
  \bibAnnoteFile{NoStop}{AL09}%
\bibitem{Williams1952ssr}%
  \BibitemOpen
  \bibfield{author}{%
  \bibinfo {author} {\bibfnamefont{W.~L.}\ \bibnamefont{Williams}},\ }%
  \bibfield{journal}{%
  \bibinfo {journal} {J. Appl. Mech.}\ }%
  \textbf{\bibinfo {volume} {18}},\ \bibinfo {pages} {320} (\bibinfo {year}
  {1951})%
  \bibAnnoteFile{NoStop}{Williams1952ssr}%
\end{thebibliography}%
\end{document}